\documentclass[%
 reprint,
 amsmath,amssymb,
 aps,
]{revtex4-2}

\usepackage{graphicx}
\usepackage{dcolumn}
\usepackage{longtable}
\usepackage{bm}

\begin{document}

\preprint{APS/123-QED}

\title{The Resonant and Normal Auger spectra of Ozone}

\author{Simone Taioli}
\email{taioli@ectstar.eu}

\affiliation{European Centre for Theoretical Studies in Nuclear Physics and Related Areas (ECT*-FBK), Trento, Italy}
\altaffiliation[Also at ]{Trento Institute for Fundamental Physics and Applications (TIFPA-INFN), Trento, Italy \& \\ Peter the Great St. Petersburg Polytechnic University, Russia}
\author{Stefano Simonucci}%
 
\affiliation{%
 School of Science and Technology, University of Camerino, Italy
}%
\altaffiliation[Also at ]{INFN, Sezione di Perugia, Italy}

\date{\today}

\begin{abstract}
In this work we outline a general method for calculating Auger spectra in molecules, which accounts for the underlying symmetry of the system. This
theory starts from Fano's formulation of the interaction between discrete and continuum states and generalizes this formalism to deal with the simultaneous presence of several intermediate quasi-bound states and several non-interacting decay channels. Our theoretical description is specifically tailored to resonant autoionization and Auger processes, and includes explicitly the incoming wave boundary conditions for the continuum states and an accurate treatment of the Coulomb repulsion. This approach is implemented and applied to the calculation of the $K-LL$ Auger and autoionization spectra of ozone, which is a $C_{2v}$ symmetric molecule, whose importance in our atmosphere to filter out radiation has been widely confirmed. 
We also show the effect that the molecular point group, and in particular the localization of the core-hole in the oxygen atoms related by symmetry operations, has on the electronic structure of the Auger states and on the spectral lineshape by comparing our results with experimental data.
\end{abstract}

\maketitle

\section{Introduction}

Ozone has attracted enormous interest in recent years, due in particular to its paramount role played in atmospheric chemistry. On the one hand, a dramatic decrease in ozone concentration within the Earth's upper atmosphere has shown a major negative impact on the effective shielding of our planet from ultraviolet sunlight, whose absence would be fatal to life. \cite{doi:10.3137/ao.460101}.
On the other hand, ozone is known to act as greenhouse gas \cite{https://doi.org/10.1029/2004GL021900,FUGLESTVEDT1999961}, thus increasing global warming, as well as to be a harmful pollutant in metropolitan areas produced by
exhaust gases of combustion processes with a considerable human health hazard potential. 

These reports on the importance of ozone in the chemical dynamics of the atmosphere, as well as in the absorption of virtually all the ultraviolet light that passes into the stratosphere, fuelled ozone research. However, while several studies focused on the ground state electronic and vibrational properties of ozone \cite{Ju_Xiang_2007,BANICHEVICH1993155,doi:10.1063/1.4821638}, little to no theoretical and experimental information about electronic excited state dynamics can be found \cite{Lopaev_2008,doi:10.1063/1.472412,doi:10.1021/acs.jpca.8b01554}. This investigation turns out to be crucial e.g. to identify the molecular species that filter out the highest--energy light, such as $\gamma$, X, and UV-rays and to determine the amount of radiation reaching the surface of our Earth.

Moreover, the analysis of the excited states is interesting also in view of the fact that all of the ozone bands are dissociative \cite{doi:10.1063/1.1385156,doi:10.1063/1.4941559}, meaning that the molecule falls apart to $O+O_2$ upon photon absorption. This dissociative mechanism drives further helpful or harmful reactions depending on where in the Earth's atmosphere they occur~\cite{doi:10.3137/ao.460104}.
Often in these photo-excited processes, the system goes through intermediate quasi-bound states, which have long lifetimes if compared with the collision times. Afterwards, the excited system undergoes a dissociative path into final channels that are characterized by the presence of a few asymptotically non-interacting fragments, the observation of
which provides useful information on the properties of the system under examination. In general, these decay processes are due to the fact that the resonant excited state produced by the initial collision with the high-energy photon is embedded in the continua of the final fragments.  

Auger spectroscopy, in particular, is a type of electron spectroscopy based on scattering processes in which the initial state consists of a projectile,
typically X-ray or UV photons or electrons, that collides with an atomic, molecular or solid state target, and the final states feature electrons, photons or heavy particles according to the type of experiment. 
The analysis of the energies and the intensity distribution of these secondary electrons represents the central problem of Auger spectroscopy.
Indeed, Auger electrons are emitted with system-dependent probabilities in several open channels, characterised by different kinetic energies and identified by different quantum numbers. Of course the Auger spectral lineshape reflects the symmetry of the perturbed system.

From the standpoint of computation one notes the existence of an ample body of
techniques developed for atomic systems, whereas for molecules the number is much
more limited \cite{taioli2009surprises,taioli2010electron}. This is mainly due to the difficulties created by the reduced symmetry of
molecular systems, which hinders the use of numerical techniques for representing the electron
as it moves outward through the field of the ionized molecule. Furthermore, another
relevant difficulty of molecules is represented by the inclusion of the nuclear motion and its coupling with the electronic degrees of freedom. For example, in ozone, fast excited dissociation may occur at once with the Auger emission \cite{doi:10.1063/1.1385156}, generating an interplay between the vibrational and electronic degrees of freedom beyond the Born-Oppenheimer approximation \cite{verucchi2012epitaxy,tatti2016synthesis,segatta2017quantum}, which affects the kinetic energy of the emitted Auger electrons. 

In this work we present first a general theoretical and computational method for interpreting core-electron
spectroscopies on molecules, such as autoionization and Auger, which allows one to reproduce accurately the experimental
results, by including in principle the vibrational details as well as the combined
effects on the spectral lineshapes due to the intrinsic features of the transition, the
characteristics of the incident radiation and those of the electron spectrometer. In particular, we discuss the central feature of our approach, that is the ability
to calculate accurate wave functions for continuum states of polycentric symmetric (or asymmetric) systems at a computational effort similar to that of standard
bound-state calculations. 
This method is used for calculating the $O1s \rightarrow \sigma^*$ resonant photoemission and the $K-LL$ Auger spectra of the
ozone molecule. The accuracy of the results is increased progressively by using different levels of theory, from mean field Hartree-Fock (HF) to correlated method, such as Configuration Interaction (CI), measuring the relative importance of
the different contributions of the electron-electron correlation to the experimental spectrum. Furthermore, the effect on the Auger lineshape of the presence of core-hole localized in the $1s$ orbitals of symmetry-related oxygen atoms rather than in a delocalized molecular orbital is discussed.

\section{Theoretical and Computational Methods}

\subsection{Autoionization and Auger decay as resonant multichannel processes}

Auger states are excited, quasi-bound states created by inner-shell excitation or ionization that are embedded in the
continuum of the next higher charge state of the system. They can decay either by
electron emission (radiationless transition) or by radiation emission. The radiationless
decay is called autoionization or Auger transition depending weather the system undergoes initially a neutral excitation by promoting an electron to an empty orbital or is ionized by the impinging photons. Often the primary excitation process is accompanied by the simultaneous ionization
(shake-off) or excitation (shake-up) of a valence electron. The resulting peaks in the
spectrum are called satellite lines. Shake-up and shake-off processes typically account for 10\% of the total spectral intensity and will be neglected in our analysis.

To evaluate the autoionization (or Auger) cross section we rely on the time-independent approach developed by Fano \cite{PhysRev.124.1866}. This approach interprets the decay process as due to the interaction between a quasi-bound state, produced by the primary excitation (ionization), and the continuum states of the single (double) ionized target.

Let us describe this method in the case of an autoionization process in which one electron is excited to a discrete (resonant) state $|\Phi>$ that is degenerate with several continua $\{|\chi_{\alpha,{\cal E}}^->\}$ of the
ionized target \cite{1982HD469A}. 
We notice that the label $|\alpha>$ is a notation standing for ``good quantum numbers'' (angular momentum, spin, etc...) in spherically symmetric systems, such as atoms, and symmetric molecules.
The total Hamiltonian of the problem, assuming the validity of the Born-Oppenheimer approximation, is 
$H=H_s+H_{rad}+H_{int}$, where $H_s$ is the many-body Hamiltonian of the isolated system, $H_{rad}=\sum_{k,\lambda}\hbar\omega a^{\dagger}_{\lambda}({\bm k})a_{\lambda}({\bm k})$ is the electromagnetic radiation Hamiltonian, where ${\bm k}$ is the wavenumber of a wave confined within a box of volume $V=L\times L \times L$, $\lambda$ is the polarization direction and $\omega=kc$, and finally the interaction operator $H_{int}=\hat e_\lambda
\cdot\sum_j q_j\vec r_j$, where $O_{\lambda}$ is the component of the dipolar operator in the
polarization direction $\hat e_\lambda$.

The continuum wave function of the system $\{|\chi_{\alpha,{\cal E}}^->\}$ is characterized by
the incoming wave boundary condition ($-$), which means that
asymptotically represents one electron released, with energy ${\cal E}$, into a channel specified by the state $\alpha$ of the ionized target at the
energy $E_\alpha$. In Fano's approach the
scattering eigenstate $|\Psi_{\alpha,{\cal E}}^->$ of the Hamiltonian of
the isolated system $H_s$ is represented by the
following linear combination of discrete and continuum states:
\begin{equation}
|\Psi^-_{\alpha,{\cal E}}>=a_\alpha({\cal E})|\Phi>+
\sum_{\beta=1}^{N_c}\int\limits_0^\infty|\chi^-_{\beta,\tau}>C_{\beta,
\alpha}(\tau,{\cal E})d\tau, \label{TF1}
\end{equation}
with the normalization condition:
\begin{equation}
<\Psi^-_{\alpha,{\cal E}}|\Psi^-_{\beta,{\cal E}^\prime}>=
\delta_{\alpha\beta}\delta(E_\alpha+{\cal E}-E_\beta-{\cal E}^\prime).
\label{TF2}
\end{equation}
The states $\{|\chi^-_{\beta,\tau}>\}$, which appear in Equation~\eqref{TF1},
are continuum states obtained through the diagonalization of the many-electron Hamiltonian matrix of the system $H_s$
that is constructed using a set of interacting continuum
states $\{|\chi_{\beta,\epsilon}\}>$ as follows:
\begin{widetext}
\begin{equation}<\chi_{\beta,\varepsilon}|H_s-E|\chi_{\gamma,\varepsilon^\prime}>=
(E_\beta+\varepsilon-E)\delta(E_\beta+\varepsilon-E_\gamma-\varepsilon^\prime)
+V_{\beta\gamma}(\varepsilon,\varepsilon^\prime;E).\label{TF3}\end{equation}
\end{widetext}
A solution of this set of equations is characterized by the
following asymptotic behavior:
\begin{equation}\lim_{r\rightarrow\infty}\chi_{\alpha,\cal E}^-=
\sum_\gamma{\Omega_\gamma\over 2ir}\left(e^{i\theta_\gamma({\cal E},r)}
\delta_{\gamma\alpha}-e^{-i\theta_\gamma({\cal E},r)}S^+_{\gamma\alpha}
\right),\label{TF4}\end{equation}
where $\{\Omega_\gamma\}$ are symmetry-adapted wave functions describing
possible states of the ionized target and containing also the angular
coordinates and the spin state of the outgoing electron.
The radial phases $\{\theta_\gamma({\cal E},r)\}$ depend on the nature
of the long-range interaction inside each channel and $S^+_{\gamma\alpha}$
is the scattering amplitude into channel $\gamma$.
The coefficients of Equation~\eqref{TF1} are obtained by solving the following equations
\begin{equation}<\Phi|H_s-E|\Psi^-_{\alpha,{\cal E}}>=<\chi^-_{\beta,\varepsilon}
|H_s-E|\Psi^-_{\alpha,{\cal E}}>=0,\label{TF5}\end{equation}
which explicitly read:
\begin{equation}(E_\Phi-E)a_\alpha({\cal E})+\sum_{\beta=1}^{N_c}\int\limits_0^\infty
M_\beta^-(\tau,E)C_{\beta\alpha}(\tau,{\cal E})d\tau=0;\label{TF6}\end{equation}
\begin{equation} a_\alpha({\cal E})M_\beta^-(\varepsilon,E)^*+(\varepsilon+
E_\beta-E)C_{\beta\alpha}(\varepsilon,{\cal E})=0, \label{TF7}\end{equation}
where
\begin{equation}M_\beta^-(\varepsilon,E)=<\Phi|H_s-E|\chi_{\beta,\varepsilon}^->;~~~~(E_\Phi-E)=<\Phi|H_s-E|\Phi>~.\label{TF8}\end{equation}
By moving into the complex plane to avoid singularities that 
appear in the coefficients 
$\{a_\alpha,C_{\beta\alpha}\}$ of Equations~\eqref{TF6},\eqref{TF7},
one gets the following expression
\begin{widetext}
\begin{equation}|\Psi^-_{\alpha,{\cal E}}>=|\chi^-_{\alpha,{\cal E}}>+
{M^-_\alpha({\cal E},E)\over E-E_r-i{\Gamma\over 2}}
\left[|\Phi>+\lim_{\nu\rightarrow 0}\sum_{\beta}
\int\limits_0^\infty{|\chi^-_{\beta,\tau}>M^-_\beta(\tau,E)^*\over
E-E_\beta-\tau-i\nu}d\tau\right],\label{TF9}\end{equation}
\end{widetext}
with $\Gamma$ and $E_r$ defined as follows
\begin{equation}\Gamma=\sum_{\beta}\Gamma_\beta=2\pi\sum_{\beta}
|M^-_\beta({\cal E}_r,E)|^2;\qquad {\cal E}_r=E_r-E_\alpha;
\label{TF10}\end{equation}
\begin{equation}E_r=E_\Phi+\Delta;\qquad\Delta=\sum_{\beta}
{\cal P}\int\limits_0^\infty{|M^-_\beta(\tau,E)|^2\over
E-E_\beta-\tau}d\tau~.\label{TF11}\end{equation}
Finally, knowing the stationary states $\{|\Psi^-_{\alpha,{\cal E}}>\}$ for each possible combination of the quantum numbers $\alpha$ and ${\cal E}$,
one can obtain the cross section for the autoionization process in which photon absorption of
appropriate energy promotes an
electron to a
discrete resonant state $|\Phi>$, and afterward the ionized target decays non-radiatively into
channel $|\chi^-_{\alpha,{\cal E}}>$.
Indeed, according to the general theory of scattering \cite{Newton:1982qc,Taylor} the autoionization cross section is proportional to the square element of the transition matrix.
Using Equations~\eqref{TF8}-\eqref{TF11} one obtains:
\begin{eqnarray}
{\bf T}^+_{fi}({\cal E})=<\chi^-_{\alpha,{\cal E}};N-1,\omega
|\hat H_{int}|o;N,\omega> +\nonumber \\+{<\chi^-_{\alpha,{\cal E}}|\hat H_s-E|\Phi>
<\Phi^-;N-1,\omega|H_{int}|o;N,\omega>
\over {\cal E}-{\cal E}_r+i{\Gamma\over 2}}\nonumber \\,\label{TF11d}
\end{eqnarray}
connecting the initial state $|\Psi^+_i>=|o>|N,\omega>$, which is the tensorial product of the
ground state $|o>$ of the system and of the $|N,\omega>$ state of
the radiation with $N$ $\omega$-frequency photons, to the final state $|\Psi^-_f>=|\Psi_{\alpha,\cal E}^->|N-1,\omega>$, which is the tensorial product
of the continuum state $|\Psi_{\alpha,\cal E}^->$ of the system and of
a state of the radiation in which one photon of frequency $\omega$ has been
absorbed. In Equation~\eqref{TF11d} the following definition has been used:
\begin{eqnarray}|\Phi^->&=&|\Phi>+\lim_{\nu\rightarrow 0}\sum_{\beta=1}^{N_c}
\int\limits_0^\infty{|\chi^-_{\beta,\tau}>M^-_\beta(\tau,E)^*\over
E-E_\beta-\tau-i\nu}d\tau; \nonumber\\ {\cal E}&=&\omega-(E_\alpha-E_o).
\label{TF11d1}
\end{eqnarray}

By neglecting the first term in the right hand side of
Equation~\eqref{TF11d}, which represents the contribution to the cross section of direct photoemission, the cross section of the resonant autoionization process can be approximated as follows:
\begin{equation}
{d\sigma_{fi}({\cal E})\over d{\cal E}}={4\pi^2\omega\over Nc}
|T^+_{fi}({\cal E})|^2=\sigma(\omega){1\over 2\pi}{\Gamma_\alpha
\over ({\cal E}-{\cal E}_r)^2+\Gamma^2/4},\label{TF11e}
\end{equation}
where $\sigma(\omega)\propto|<\Phi;N-1,\omega|H_{int}|o;N,\omega>|^2$
is the cross section of the excitation process $|o>\rightarrow|\Phi>$.
One can observe that, as long as $\Gamma_\alpha$, which represents the
decay rate into channel $|\alpha>$ and $\Gamma$, which represents the total
decay rate, are
slowly energy dependent in the range of interest,
the cross section per unit energy defined by
Equation~\eqref{TF11e} has a Lorentzian profile.
The energy shift
$\Delta$, defined in Equation~\eqref{TF11} usually represents a small
correction to $E_\Phi$, so typically can be neglected. 

We notice that similar results can be obtained also for the Auger decay where, at odds with the resonant autoionzation process, the intermediate state is
constituted by a quasi-bound state of the ionized target plus one electron emitted upon the initial photoionization, while the final state is characterized by the presence of two electrons in the continuum. In our model, which includes explicitly the incoming wave
boundary conditions for the continuum states, the Auger decay is described as a two-step process in which no interaction occurs between the primary photo-emitted and Auger electrons. 

The ${\bf T}$ matrix of the process can be, therefore, written as follows:
\begin{widetext}
\begin{equation}
T^{+\prime}_{\alpha i}=
<\chi^-_{\alpha,{\cal E}_1{\cal E}_2};N-1,\omega|\hat H_{int}
|o;N,\omega>  +\int\limits_0^\infty{<\chi^-_{\alpha,{\cal E}_1{\cal E}_2}|
\hat H_s-E|\Phi_\tau><\Phi_\tau;N-1,\omega|H_{int}|o;N,\omega>d\tau\over
{\cal E}_1+{\cal E}_2-{\cal E}_r-\tau+i\Gamma/2},\label{TF11h}
\end{equation}
\end{widetext}
where ${\cal E}_1$ and ${\cal E}_2$ are the energies of the outgoing
electrons related to the other
characteristic energies of the problem by the following relationship:
${\cal E}_1+{\cal E}_2=\omega-(E_\alpha-E_o)$ ($E_o$ is the ground state energy of the system). The intermediate
state $|\Phi_\tau>$ is a continuum state that, as long as
$\omega>>E_\alpha-E_r$, can be approximated by an antisymmetrized product,
$\hat A\{|\Phi>|\tau>\}$, where $|\tau>$ represents the state of the
primary electron.
In a similar way, the final state can be represented by
$\hat A\{|\chi^-_{\beta,{\cal E}_2}>|{\cal E}_1>\}$, where
$|{\cal E}_1>$ is the state of the primary electron. 

Neglecting the first terms in the right-hand side of
Equation~\eqref{TF11h}, that represents the probability amplitude
of the double direct ionization process, one gets
\begin{eqnarray} {d\sigma_\alpha({\cal E})\over d{\cal E}}=\sigma(\overline{\cal E})
{1\over 2\pi}{\Gamma_\alpha\over ({\cal E}-{\cal E}_r)^2+
\Gamma^2/4}; \nonumber \\ \overline{\cal E}=\omega-(E_r-E_o),
\label{TF11l}\end{eqnarray}
where $\sigma(\overline{\cal E})$ is the photoionization cross section of
the primary process. This result, again, displays a Lorentzian behaviour and indicates that
$\Gamma_\alpha$ is the rate of the non-radiative decay process
from the ion intermediate state $|\Phi>$ into the channel $|\alpha>$, and $\Gamma$ is the total Auger decay rate.

The cross sections~\eqref{TF11e}, \eqref{TF11l} represent the starting point
for constructing the ``theoretical'' spectrum to be compared with the experimental
one. They give the contributions to the spectrum due to the intrinsic features of the
target, while the finite resolution of the electron spectrometer, the specific characteristics of the incident photon beam, and the broadening of the Auger lineshapes due to vibrational modes can be taken into account by performing a convolution with a Gaussian function \cite{taioli2010electron}. 

\subsection{The projected potential approach and the many-body Hamiltonian}

Basically, to assess the cross sections~\eqref{TF11e}, \eqref{TF11l} we are left with the problems of calculating the intermediate resonant state and the continuum wavefunction. In particular, the problem of finding the eigensolutions of the Hamiltonian is separated with respect to finding the bound and continuum orbitals. This means to neglect the effects that the continuum orbital has on the bound
orbitals, which is obtained without taking into account the presence of the outgoing electron. This is the so-called {\it ''Static Exchange Approximation''}. In general, the interacting decay channels $\{\tilde\chi_{\alpha\vec k}\}$ are represented by the tensor product 
\begin{equation}
\chi_{\alpha\vec k}(1,..N-1,N)=\hat {\cal A}
[\Theta_{\alpha}(1,...,N-1)\varphi_{\alpha\vec k}(N)]\label{TH5}
\end{equation}
between a set of functions $\varphi_{\alpha\vec k}(1)$ that describes the spin-orbital of the unbound electron and $\Theta_{\alpha}(1,...,N-1)$ that is the determinant representing the bound state of the remaining $N-1$ system ($N-2$ for Auger processes), and $\hat {\cal A}$ is the antisymmetrizer that includes 
also the normalization constant. 

On the one hand, since both the intermediate
and the final ionic states are bound states of the molecule, one can
calculate their energies using standard quantum
mechanical techniques proposed for bound-state calculations, such as the single Slater determinant HF or many configuration wave functions methods, such as CI or
MC-SCF, or Green's function based techniques \cite{taioli2010electron}. 
In particular, the solutions of the secular problem for the
$(N-1)$-electron system ($N-2$ for Auger processes) are obtained inside the space spanned by the orbitals taken from
a set of $n$ bound orbitals $|\theta_j>$.
Both the HF single-determinant wavefunction and the CI treatment of symmetry-adapted orbitals will be used to obtain the bound state eigenfunctions in Equation~\eqref{TH5} as:
\begin{equation}
<\Theta_{\alpha}|\hat H_s^{N-1}|\Theta_{\beta}>=
E_{\alpha}\delta_{\alpha\beta}.\label{TH6}    
\end{equation}
This solution is searched within the Hilbert space spanned by a basis set that we use to expand the orbital wavefunctions.
Typically our basis set is built with Hermite Gaussian functions (HGF)
\begin{widetext}
$$g(\mathbf{r})=g(u,v,w;a,\mathbf{R};\mathbf{r})=N{\partial^{u+v+w}\over
\partial X^u \partial Y^v\partial Z^w}({2\alpha\over\pi})^{3/4}
exp[-\alpha (\mathbf{r}-\mathbf{R})^2],$$
\end{widetext}
where $\mathbf{R}\equiv(X,Y,Z)$ gives the position where $g$ is centered, $\alpha$ is a coefficient determining the HGF width, the order of derivation ($u,v,w$) determines the symmetry type, and $N$ is a
normalization factor
$$N=[\alpha^l(2u-1)!!(2v-1)!!(2w-1)!!]^{-1/2}{,}\qquad l=u+v+w,$$
Using a mixed basis set of HGFs of every order and centrature we evaluate the mono and bi-electronic integrals of the many-body Hamiltonian~\eqref{TH6}. 

On the other hand, to construct the continuum orbital of Equation~\eqref{TH5} in the effective field of the bound orbitals we developed a method that is capable to include also the
interchannel coupling among various $\{\chi_{\alpha\vec k}\}$ and,
furthermore, the interaction among continuum and discrete states. In order
to include these effects we use the projected
potential approach. Within the projected
potential framework,
a model Hamiltonian is defined in which both the
monoelectronic and the bielectronic part of the potential are
represented in terms of $L^2$-functions.
The use of this model Hamiltonian allows us to include also the interchannel 
coupling and to obtain directly the correct, non-interacting decay 
channels $\{|\chi^-_{\alpha\vec k}>\}$ defined in the previous section (see Equation~\eqref{TF4}).

To present this approach we consider the case of a $N$-electron system with one electron in the continuum. We define a projector $\hat\pi(i)=\sum_{l=1}^{m}~|g_l(i)><g_l(i)|$ into a $m$-dimensional space
$({\cal G})$ of $L^2({\cal R}^3)-$functions, spanned by  the orthonormal set $\{|g_l>~;~l=1,..,m\}$.
We project both the electron-nuclei attraction potential $\hat V^{en}(i)=-\sum_{\mu}{Z_{\mu}\over|\vec r_i-\vec
R_{\mu}|}$, and the electron-electron Coulomb repulsion $\hat v(i,j)=
{1\over|\vec r_i-\vec r_j|}$
operators, obtaining the following model electronic Hamiltonian 
\begin{equation}
\hat H_s(1,..,N)=\sum^N_{i=1}[\hat T(i)+\hat V_{en}^{\pi}(i)]
+{1\over 2}\sum^N_{i\neq j}\hat v^{\pi}(i,j),\label{TH1}
\end{equation}
\begin{equation}
\hat T(i)=-{1\over 2}\nabla^2_i~~~;~~~
\hat V_{en}^{\pi}(i)=\hat\pi(i)\hat V_{en}(i)\hat\pi(i),\label{TH2}\end{equation}
\begin{equation}\hat v^{\pi}(i,j)=\hat\pi(i)\hat\pi(j)\hat v(i,j)\hat\pi(i)\hat\pi(j).
\label{TH3}
\end{equation}

By projecting out only the potential terms of the Hamiltonian we come up with the solution to the issues of i) representing accurately the orbital inside the scattering region, which is the important volume where the matrix elements that
couple bound and continuum states (see Equation~\eqref{TF8}) have to be appropriately calculated; ii) recovering the continuum part of the spectrum by means of the unprojected kinetic term. The region where the potential is projected can of course be incremented by including e.g. more diffuse HGFs so to approximate more accurately the Coulomb potential tail.
Outside this region, where only the long-range part of the potential survives, one can represent
the continuum orbital as a linear combination of eigenfunctions of the long-range potential, which are typically known analytically (e.g. Coulomb wave functions). 

In order to construct the product~\eqref{TH5} and find the continuum eigenfunctions $\varphi_{\alpha\bf k}$ of the Hamiltonian~\eqref{TH1} with positive eigenvalues, we notice that the Hamiltonian matrix diagonalization within the entire functional space $\cal G$ is equivalent to solve the following projected Lippmann-Schwinger (LS) equation, that includes also the appropriate boundary condition:
\begin{equation}
\varphi_{\alpha {\bf k}}(\bm r)=e^{i\bf k\cdot\bf r}+
\hat G_0^-(\varepsilon)\hat T_\alpha(\varepsilon)e^{i\bf k\cdot\bf r},
\label{ML11}\end{equation}
where $G_0^-(\epsilon=\frac{k^2}{2})=\frac{1}{\epsilon-\frac{\hat k^2}{2}-i\epsilon}$ is the free-particle Green's function and $\hat T_\alpha$ is the transition operator defined by the equation:
\begin{equation}
\hat T_\alpha=\hat V^{\pi}_\alpha+\hat V^\pi_\alpha\hat G_0^-(\varepsilon)
\hat T_\alpha.\label{ML12}
\end{equation}
In Equation~\eqref{ML12} $\hat V^\pi_\alpha(\bm r)$ is the approximate representation of the Coulomb operator (see Equation~\eqref{TH2}), including both electron-electron and electron-nuclei interactions, projected by a finite set of $L^2$ functions. The elements of this basis set are chosen to minimize the difference
$(\hat V_\alpha-\hat V^\pi_\alpha)|\varphi_{\alpha\vec k}>$ inside the scattering volume. We notice that in Equation~\eqref{ML11} the correction to the free-wave state to obtain the scattering wavefunction is represented by a linear combination of functions of the type 
$G_0^- |\alpha \rangle g_j$.

By using the previous definitions
one gets the following expression for the matrix element that couples
two interacting channels:
\begin{widetext}

\begin{equation}<\chi_{\alpha\vec k}|\hat H^N|\chi_{\beta\vec p}>=
(2\pi)^3\delta(\vec k - \vec p)\delta_{\alpha\beta}({k^2\over2}+E_{\alpha})+
<\varphi_{\alpha\vec k}|\hat V^{en}_{\pi}\delta_{\alpha\beta}+
\hat W_{\pi}^{\alpha\beta}|\varphi_{\beta\vec p}>,\label{TH11}\end{equation} 
\end{widetext}
where

\begin{widetext}
\begin{equation}
\hat W_{\pi}^{\alpha\beta}(1)=\sum_{j=2}^N
<\Theta_{\alpha}(2,.j.,N)|\hat v_{\pi}(1,j)(\hat 1- \hat {\cal P}_{1,j})|
\Theta_{\beta}(2,.j.,N)>,\label{TH12}\end{equation}
\end{widetext}
and $\hat {\cal P}_{1,j}$ is the operator that interchanges the
$(1,j)$ variables. 

If now we assume that the bound-state problem has been
solved, i.e. the eigenvectors $(|\Theta_1>,|\Theta_2>,..,|\Theta_M>)$ of
$\hat H^{N-1}$ ($\hat H^{N-2}$ for the Auger process) have been found inside the space of the Slater determinants
built up using the $\{\theta_j\}$ orbitals, we
can look at the matrix elements defined in Equation~\eqref{TH11} as the
representation of an effective one-particle Hamiltonian over a set of
basis vectors \hbox{$\{|\varphi_{\alpha\vec k}>\}$}. This Hamiltonian, indeed, is that of
a particle, with internal degrees of freedom, that moves in an effective
potential depending on the internal states $\{|\alpha>\}$ of the particle
itself.
We notice that the basis vectors $\{|\varphi_{\alpha\vec k}>\}$ are labelled by two
indices, one continuous $(\vec k)$ and one discrete $(\alpha)$, and that they shall satisfy
orthonormality constraints both with respect to $\vec k$ and to $\alpha$. 
Finally, the bound orbitals $\Theta_{\alpha}(1,...,N-1)$ and continuum wavefunctions $\varphi_{\alpha\vec k}(\vec r)$ can be made mutually orthogonal.

\section{Results and discussion}

\subsection{The ozone molecular geometry and its electronic structure}

\begin{figure}[hbt!]
\includegraphics[trim=4cm 4cm 4cm 4cm,clip,angle=-90,width=0.9\linewidth]{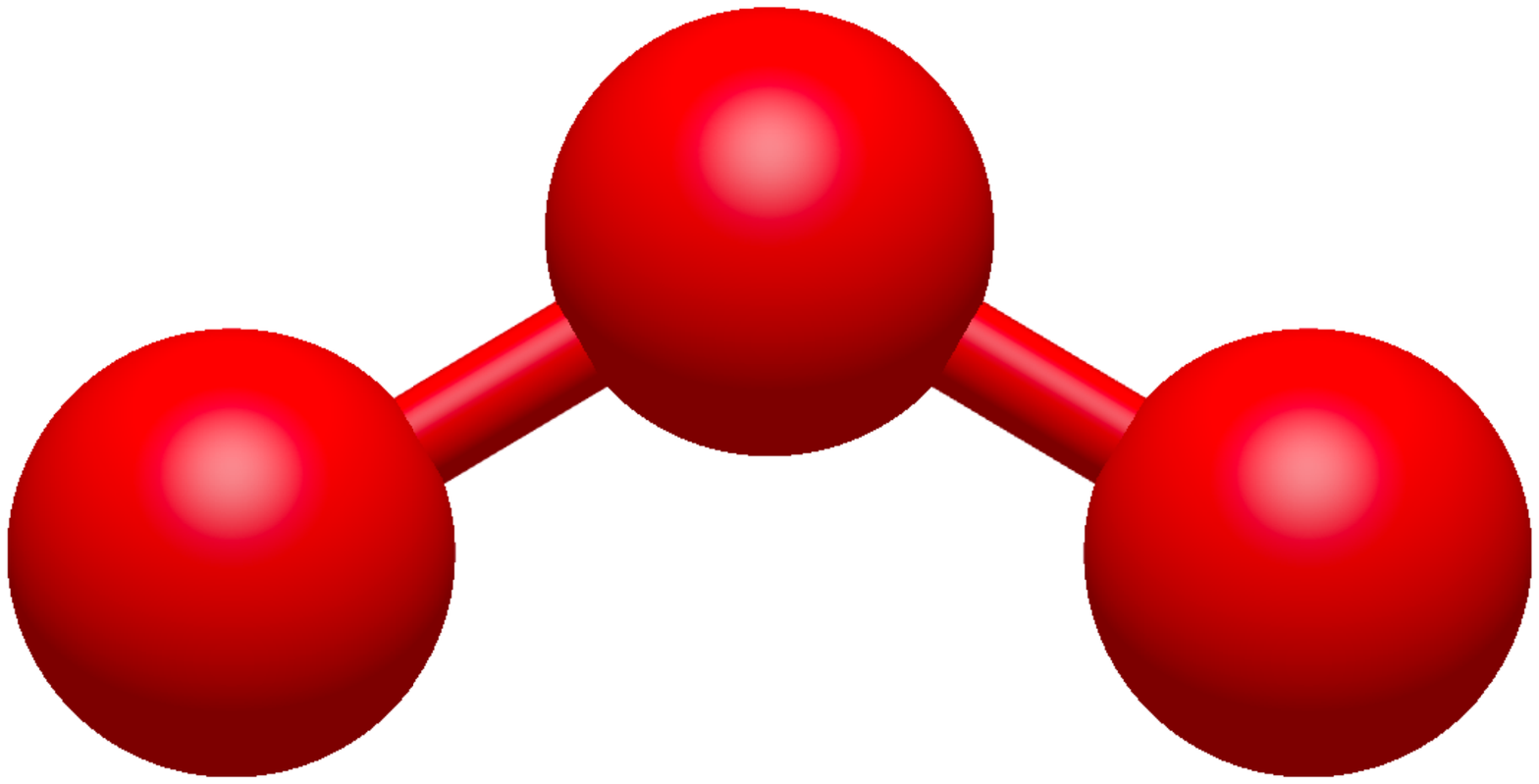}
\caption{Ozone geometry in the ground state. \label{fig1}}
\end{figure}

Ozone ground state has a trigonal planar bent molecular geometry belonging to the $C_{2v}$ symmetry group (similar to the water molecule), whereby the central oxygen atom is in a $sp^2$-hybridized configuration.
To optimize the ozone atomic coordinates we started from experimental oxygen positions and accommodated the molecule in a cell with side of 10 \AA. The ozone geometry was relaxed 
below $10^{-3}$ Ry/\AA{} for the interatomic forces via first-principles density functional theory (DFT) calculations  as implemented in the Quantum Espresso code suite \cite{Giannozzi_2009}, using a PBE-GGA functional \cite{PBE}.
We have used the Troullier-Martins (TM) norm-conserving pseudopotentials tabulated in the Quantum Espresso web page. Including the $\Gamma$ point only to sample the Brillouin zone and using a kinetic energy cut-off of 130 Ry, the self-consistent DFT convergence is reached within the energy error of $10^{-5}$. 
Upon optimization, the O--O bond length turns out to be 1.273 \AA, while the O--O--O angle is 117.16$^\circ$, which well compare with experimental data \cite{TANAKA1970538} (1.272 \AA~and 116.78$^\circ$, respectively). The ozone ground state geometry is shown in Figure~\ref{fig1}.
The molecular bonds in  ozone can be represented as a resonance between two contributing structures, each with a single bond on one side and double bond on the other, where
the terminal O$_T$ atoms are more electron rich than the central O$_C$ atom.  
The 3 $sp^2$ hybrid orbitals form a net of 2 O--O $\sigma$ bonds and 
5 lone pairs (two on each terminal O$_T$ and one on the central O$_C$). The remaining 4 valence electrons are distributed among the 3 unhybridized $2p_z$ orbitals on each oxygen atom, forming one lowest energy $\pi$ bond, one highest energy $\pi^*$ antibonding and one intermediate energy $\pi$ nonbonding orbitals, respectively.

\subsection{The Auger spectrum of ozone}

Electronic structure calculations have been carried out initially at mean-field HF level for the ground state, a set of highly-excited core-hole intermediate states and the double-ionized molecule. 
We remind that the HF method is based on a set of non-linear equations, minimized to deliver the best variational wavefuntions and energy. Thus, they are sensitive to the kick-off orbitals. In this regard, we will show below the effect of localizing the inner hole on atomic or molecular symmetry orbitals.  

The ground state of the
ozone molecule is described by the following configuration
$^1A_1:~ 1a_{1}^{2}\,1b_{1}^{2}\,2a_{1}^{2}\,3a_{1}^{2}\,2b_{1}^{2}\,4a_{1}^{2}\,5a_{1}^{2}\,3b_{1}^{2}\,1b_{2}^{2}\,4b_{1}^{2}\,6a_{1}^{2}\,1a_{2}^{2}$.

The intermediate quasi-bound states populated by the primary ionization are the following:
\begin{itemize}
\item	$^2A_1:~ 1a_{1}^{1}\,1b_{1}^{2}\,2a_{1}^{2}\,3a_{1}^{2}\,2b_{1}^{2}\,4a_{1}^{2}\,5a_{1}^{2}\,3b_{1}^{2}\,1b_{2}^{2}\,4b_{1}^{2}\,6a_{1}^{2}\,1a_{2}^{2}$;
\item	$^2B_1:~ 1a_{1}^{2}\,1b_{1}^{1}\,2a_{1}^{2}\,3a_{1}^{2}\,2b_{1}^{2}\,4a_{1}^{2}\,5a_{1}^{2}\,3b_{1}^{2}\,1b_{2}^{2}\,4b_{1}^{2}\,6a_{1}^{2}\,1a_{2}^{2}$;
\item $^2A_1:~ 1a_{1}^{2}\,1b_{1}^{2}\,2a_{1}^{1}\,3a_{1}^{2}\,2b_{1}^{2}\,4a_{1}^{2}\,5a_{1}^{2}\,3b_{1}^{2}\,1b_{2}^{2}\,4b_{1}^{2}\,6a_{1}^{2}\,1a_{2}^{2}$.
\end{itemize}

The final states of the doubly ionized target
are characterized by two holes distributed in all the possible ways among the 3-6$a_1$, 2-4$b_1$, 1$a_2$, and 1$b_2$ orbitals, which results in 81 different possible final channels.  
We will limit the discussion to the
electronic spectrum, using only the electronic Hamiltonian and taking into account
the effects of the nuclear motion in a simplified manner by adding a broadening of the Auger lines only at the end, in order to compare the theoretical and experimental spectra. A further line broadening will be included to take into account the finite resolution of the spectrometer used in electron spectroscopy measurements. Furthermore, the two-step
and static exchange approximations are used and only the decay process from the intermediate quasi-bound
state is analyzed. In the case of the Auger decay, we further assume that the primary electron is fast enough to avoid any appreciable interaction with the ionized ozone molecule.
\begingroup
\squeezetable
\begin{table}[hbt!]
\begin{tabular}{ccccccc}
\toprule
 $~~~~~$ & \textbf{$1a_{1}$} & \textbf{$1b_{1}$}	& \textbf{$2a_{1}$} & \textbf{$3a_{1}$} 	& \textbf{$2b_{1}$}	& 	 \textbf{$4a_{1}$}	 \\
\hline
$\epsilon_{HF}$ non rel. & -20.9179 & -20.7070 & -20.7070 & -1.7453 & -1.4281 & -1.0948 \\
$\epsilon_{HF}$ rel.  & -20.9124 & -20.7015 & -20.7015 & -1.7460 & -1.4286 & -1.0957 \\

\toprule
$~~~~~$ &\textbf{$5a_{1}$}	& 	 \textbf{$3b_{1}$}	& \textbf{$1b_{2}$} & \textbf{$4b_{1}$} & \textbf{$6a_{1}$} &  \textbf{$1a_{2}$} \\
\hline
$\epsilon_{HF}$  non rel. & -0.8301 & -0.7983 & -0.7785 & -0.5640 & -0.5531 & -0.4870 \\
$\epsilon_{HF}$ rel. & -0.8296 & -0.7979 & -0.7778 & -0.5633 & -0.552 & -0.4865\\
\hline
$E_{HF}$ non rel. & -224.356 \\ 
$E_{HF}$ rel.     &  -224.408 \\
\end{tabular}
\caption{Relativistic (rel.) vs. non relativistic (non rel.) HF total ($E_{HF}$) and orbital energies ($\epsilon_{HF}$), calculated for the ozone ground state. Values are reported in a.u.\label{tab1}}

\end{table}\endgroup

The basis functions used in the expansion of bound and continuum orbitals is taken from the aug--cc--pVQZ basis set with $(12s,6p,3d,2f,1g)$ HGF contracted to $(5s,4p,3d,2f,1g)$ centered on each oxygen nucleus. To increase the accuracy and rely on a larger number of diffuse functions to represent the continuum orbital, in our simulations we completely decontracted the aug--cc--pVQZ HGF basis set to $(13s,7p,4d)$ HGF without the $g$-symmetry Gaussian to limit the computational cost. Indeed, this decontraction procedure will add significantly to the computational cost of the calculation of the Auger spectra. The HGF basis set has been decontracted in order to have eigenvectors of an average HF operator at energies near to those of the Auger
electron in the various decay channels. Using this HGF basis set, we calculated the total and orbital energies of the neutral ground state of ozone by numerical solution of the HF equations with and without the inclusion of relativistic effects \cite{morresi2018nuclear}, where in the latter simulation the electronic repulsion takes into account also the Gaunt term \cite{reiher2009}. These results are reported in Table \eqref{tab1} in a.u. Analysing these values we conclude 
that relativistic effects are negligible and a non-relativistic approximation can be adopted to simulate the Auger lineshape of ozone.

The final states produced by the decay process are represented by the wave functions of Equation~\eqref{TH5}, where $| \Theta_\alpha \rangle $ is the wave function for the state $| \alpha \rangle$ representing the final single and double ionized target, respectively, while 
$\varphi_{\alpha\kappa}$ is the spin-orbital describing the Auger electron. We remind that the use of $\chi_{\alpha\vec k}$ instead of the well behaving wave function $\chi_{\alpha\vec k}^-$ that diagonalizes the Hamiltonian \eqref{TF3} is equivalent to disregard the coupling among the decay channels. This approximation is in general too drastic \cite{taioli2009surprises,taioli2010electron} and will be abandoned in our calculations. Intermediate and final bound states of the single and double charged molecule are
represented by the following linear combinations of Slater determinants:

\begin{widetext}
\begin{eqnarray}
|\Phi(1,\dots, N-1) \rangle &=& \frac{1}{\sqrt(N-1)!}\sum_i{a_i|\varphi_{1}^i(1),\dots,\varphi_{N-1}^i(N-1)|},\\
|\Theta_\alpha (1,\dots, N-1)\rangle &=& \frac{1}{\sqrt(N-2)!}
\sum_j{b_{\alpha j}|\theta_{1,\alpha}^j(1),\dots,\theta_{N-2,\alpha}^j(N-2)|}.
\end{eqnarray}

\end{widetext}
The coefficients $a_i$ and $b_{\alpha,j}$ are determined variationally by solving the secular problem
with respect to the standard electronic Hamiltonian. The bound orbitals $\{ \varphi \}$ and $\{\theta \}$ are obtained by solving separate HF equations for the various states of interest and therefore the resulting orbitals for a given state are not orthogonal to those of a different state of charge. 

The matrix elements (\ref{TF8}) that couple intermediate and final states are calculated between Slater determinants
built up in terms of orbitals belonging to mutually non-orthogonal sets as follows:

\begin{widetext}

\begin{eqnarray}
M_{\alpha,\vec k}(E)=<\Phi|\hat H-E|\chi_{\alpha,\vec k}>=\sum_i\sum_j a_i b_{\alpha,j}\left\{\sum_{l=1}^{N-1}
\left(\sum_{m=2}^{N-1}<\varphi^i_l|\hat h|\vartheta^j_{m,\alpha}>
|S_\alpha^{ij}(l;m)|\right. \right. \nonumber \\
+\left. <\varphi^i_l|\hat h|\eta_{\alpha,\vec k}>
|S_\alpha^{ij}(l;1)|\right)+\sum_{l=2}^{N-1}\sum_{r=1}^{l-1}
\left[\sum_{m=3}^{N-1}\sum_{s=2}^{m-1}\left(
<\varphi^i_l,\varphi^i_r|\hat g|\vartheta^j_{m,\alpha},\vartheta^j_{s,
\alpha}>\right. \right.  \nonumber \\
\left. -<\varphi^i_l,\varphi^i_r|\hat g|\vartheta^j_{s,\alpha},
\vartheta^j_{m,\alpha}>\right)|S_\alpha^{ij}(l,r;m,s)|+\sum_{m=2}^{N-1}
\left(<\varphi^i_l,\varphi^i_r|\hat g|\vartheta^j_{m,\alpha},
\eta_{\alpha,\vec k}>\right. \nonumber \\
	-<\varphi^i_l,\varphi^i_r|\hat g|\eta_{\alpha,\vec k}\vartheta^j_{m,\alpha}>)|S_\alpha^{ij}(l,r;m,1)|]-E|S^{ij}_\alpha|\},\label{M6} 
\end{eqnarray}
\end{widetext}

where $S_{\alpha}^{ij}$ is the following overlap matrix
\begin{widetext}

\begin{eqnarray}
S^{ij}_\alpha=\begin{pmatrix} <\varphi^i_1|\vartheta^j_{1,\alpha}>&\ldots&
<\varphi^i_1|\vartheta^j_{N-2,\alpha}> &<\varphi^i_1|\eta_{\alpha,\vec k}>
\cr \vdots & & \vdots&\vdots\cr
<\varphi^i_{N-1}|\vartheta^j_{1,\alpha}>&\ldots& <\varphi^i_{N-1}|
\vartheta^j_{N-2,\alpha}> &<\varphi^i_{N-1}|\eta_{\alpha,\vec k}> \end{pmatrix}
\label{M7}
\end{eqnarray}
\end{widetext}

and $|S^{ij}_{\alpha}(l;m)|$ is the determinant of the minor obtained by taking away row $l$ and
column $m$ and $|S^{ij}_{\alpha}(l, r;m, s)|$ that of the minor obtained by taking away rows $(l,r)$ and columns $(m,s)$.
In order to calculate these matrix elements, which give the
relative decay rates onto the various channels, one has to evaluate integrals between $L^2$-functions, used to represent both the bound and the continuum orbitals. Analytical expressions of these integrals using HGF of any order and centre as basis functions for the bound orbitals can be derived \cite{taioli2010electron}, which decrease the computational cost. 

We stress that, besides a truncated multi-configuration expansion, the only approximation in our bound state calculations is represented by the decoupling of the electronic and nuclear motion, without including the effects of the latter from first-principles. Therefore,
both the intermediate and the final states are purely electronic states obtained at the
equilibrium geometry of the molecule.

\begingroup
\squeezetable
\begin{table}[hbt!] 

\begin{tabular}{ccccc}
\hline
\textbf{Auger state}	& \textbf{$E_{1}$}	& \textbf{$E_{2}$} & \textbf{$E_{3}$} \\
\hline
$1a_{1}^{-1}$	& -203.705 (562.0)		& -204.168(549.4) & -204.268(546.6)\\
$1b_{1}^{-1}$	& -203.853(557.9)		& -204.373(543.8) &  -204.482(540.8) \\
$2a_{1}^{-1}$   & -203.854(557.9)   & 
-204.373(543.8) & -204.482(540.8) \\
\end{tabular}
\caption{Auger state energies in a.u. for different core-hole configurations. $E_{1}$
is the single-configuration HF energy; $E_{2}$ is calculated by using single-excitation CI (CIS) with an active space up to the 27th excited state with respect to the relevant configuration. In $E_{3}$ active space was increased to include up to 63 molecular orbitals. 
Ionization energies (in eV) at the same level of theory are also reported in parenthesis for each Auger state.
\label{tab2}}
\end{table} \endgroup
In Table~\ref{tab2} we report the HF energy $E_{1}$, calculated by assuming a single determinant wavefunction with a core-hole in the Auger states $1a_{1}^{-1}$, $1b_{1}^{-1}$ and $2a_{1}^{-1}$, respectively. Moreover, we also report in Table~\ref{tab2} the energies calculated using CIS from the reference configuration for an active space of 27 and 63 orbitals, labelled $E_{2}$ and $E_{3}$, respectively. Both $E_{1}$ and $E_{2}$ were calculated preserving the molecular symmetry, whereby the hole is not localized on a particular oxygen atom rather ``delocalized'' on a molecular symmetry orbital.
Furthermore, in Table~\ref{tab2} we report also the ionization energies (IE) necessary to extract inner electrons. We notice that IEs from $1b_{1}$ and $2a_{1}$ orbitals are almost identical, being these molecular orbitals almost localized on the two C$_{2v}$ symmetry-related oxygen atoms. While HF orbital optimization was achieved starting from a hole initially delocalized on the $1a_{1}, 1b_{1}, 2a_1$ inner shells, a partial relocalization of the holes is achieved by adding dynamic correlation via CIS. 

However, the most difficult part is the treatment of the continuum wavefunction. Experience tells us that a major
source of error is due to poor representation of the continuum orbital, which is typically expanded using plane-waves. Our approach overcome this problem by the use of the model Hamiltonian with projected potential defined in Equation~\eqref{TH1}. The construction of its scattering stationary states inside the space spanned by the orthonormal wave functions $\{\tilde\chi_{\alpha\vec k}\}$, defined in Equation~\eqref{TH5}, is obtained by the numerical solution of the LS equation \eqref{ML11}.
The analytic expressions of the elements of the matrix representative of the transition operator \eqref{ML12}  using HGF can be obtained from the knowledge of those relative to $V_{\alpha}^\pi$ and $\hat{G}_0$. 
To improve the theoretical reproduction
of the experimental Auger spectra, we finally diagonalize the interchannel Hamiltonian~\eqref{TH11}, which includes the coupling among the various decay channels, each one described by a wave function that takes into account, at CIS level of theory, the intrachannel correlation effects. The diagonalization of the interchannel coupling between interacting channels redistributes the decay probability among independent channels.

\begin{figure}[hbt!]
\includegraphics[width=0.9\linewidth]{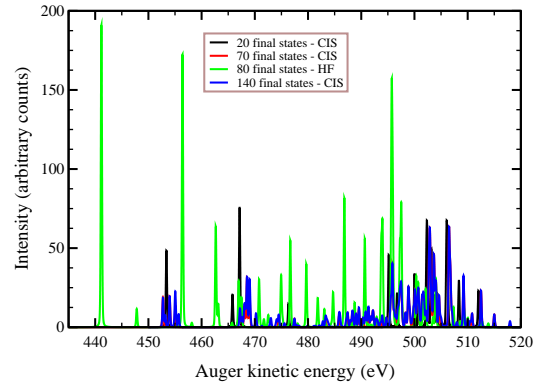}
\caption{$K-LL$ Auger spectrum of ozone, where $K=1a_1$. Green line: HF level of theory with 80 independent final states, obtained by diagonalization of the multichannel Hamiltonian~\eqref{TH11}. Black, red, and blue lines represent the spectral lineshape obtained with an active space of 27 molecular orbitals and a number of 20, 70, 140 independent final channels, respectively. \label{fig2}}
\end{figure}

In Figures~\eqref{fig2}, \eqref{fig3}, and \eqref{fig4} we plot the $K-LL$ Auger spectra of ozone calculated with the
decoupled channel approach in which the intermediate states are characterized by single inner-shell vacancies in the $K=1a_1,1b_1,2a_1$ molecular orbitals and the final states by two vacancies in all possible valence orbitals. In particular, the green line represents 80 principal decay channels using HF level of theory, while black, red and blue lines represent simulations with 20, 70, and 140 final states, respectively, using CIS level of theory. In these calculations the active space consists of 27 molecular orbitals. In Table~\eqref{tab7} we report the total $K-LL$ Auger decay rates from the $1a_1, 1b_1, 2a_1$ core-hole intermediate states as a function of the number of final states included in the calculations. 
The calculated transition energies, partial and total decay rates of the $K-LL$ Auger processes are reported in Tables~\eqref{tab4} ($K=1a_1$), \eqref{tab5} ($K=1b_1$), \eqref{tab6} ($K=2a_1$), respectively, along with the final state orbital occupations for the 20 brightest final states. 
We notice that the $K-LL$ Auger transition spectra in Figures~\eqref{fig3} and \eqref{fig4} are almost identical, and also the total Auger probability does not differ significantly, being 2.9127 $\times 10^{-3}$ a.u. and 2.9485 $\times 10^{-3}$ a.u. (see Table~\eqref{tab7}, third column), respectively. 
This result is obtained despite the symmetry of the intermediate state is different, as the initial hole created upon photoionization belongs to different irriducible representation of the $C_{2v}$ symmetry group. This is interpreted as the effect of the inclusion of the single excitations from the reference state in the description of the orbital wavefunctions, which results in the quasi-relocalization of the core-hole into one of the two symmetry-equivalent oxygen atoms despite the HF orbitals were optimized starting from a hole ``delocalized'' in each of the molecular orbitals. The gain in energy of this hole relocalization via multi-configuration interaction is about 17 eV. At variance, the total $K-LL$ Auger probability for an initial hole in the deepest $K=1a_1$ molecular orbital is lower independently of the number of final channels used (see Table~\eqref{tab7}).
We notice that the use of CIS to treat the orbital wavefunctions has large impact on the accuracy of the Auger decay rate and peak position, showing a large blueshift with respect to the HF values in excess of 10 (see Figure~\eqref{fig2}) to 20 eV (see Figures~\eqref{fig3} and~\eqref{fig4}).

The Auger lines, peaked at the transition energies calculated as by Equation~\eqref{TF10}, were broadened by Lorentzian functions, whose widths were obtained from Equation~\eqref{TF10}. Finally, to take into account the finite resolution of the experimental set-up and the vibrational broadening, we convoluted the theoretical lineshape with a Gaussian having full width at half maximum (FWHM) of $0.1$ eV. 

\begin{table}[hbt!] 

\begin{tabular}{c|ccc}
\hline
 &  & \textbf{\# of channels} & $~$ \\
\textbf{Auger state} & \textbf{20}	& \textbf{70} & \textbf{140} \\
\hline
$1a_{1}^{-1}$ & $1.4321$ & $2.1208$ & $2.6241$ \\
$1b_{1}^{-1}$ & $1.9824$ & $2.4548$ & $2.9127$ \\
$2a_{1}^{-1}$  & $1.9957$ & $2.4644$ & $2.9485$ \\
\end{tabular}
\caption{Total $K-LL$ Auger decay rates of the three different intermediate states $1a_1, 1b_1, 2a_1$ using CIS with an active space that includes up to 27 molecular orbitals for 20, 70, and 140 final channels.
Data are reported in $\times 10^{3}$  a.u.
\label{tab7}}
\end{table}

We observe that by increasing the number of non interacting final channels,
from 20 (black lines) to 70 (red lines) and 140 (blue lines), the partial
probability of Auger decay of course redistributes among different channels,
and also the total probability changes significatively, moving e.g. from
1.4321 $\times 10^{-3}$ a.u. to 2.6241 $\times 10^{-3}$ a.u. for the $K=1a^{-1}$
Auger state (see Table~\eqref{tab7}).

\begin{figure}[hbt!]
\includegraphics[width=0.9\linewidth]{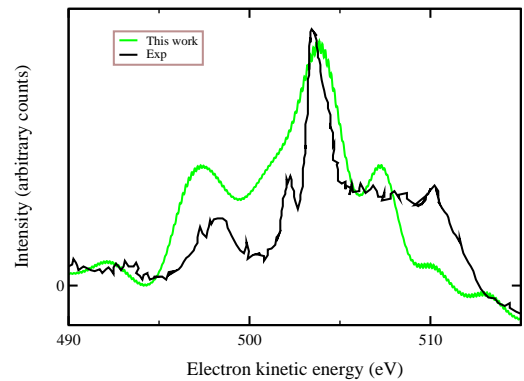}
\caption{Comparison between the $K-LL$ ($K=1a_1$) Auger experimental spectrum of ozone (black line) \cite{doi:10.1063/1.1385156} and our first-principles simulation (green line) obtained with an active space of 27 molecular orbitals and 140 independent final channels. Our lineshapes were convoluted via 0.8 eV Gaussian function to achieve the experimental broadening. \label{fig2a}}
\end{figure}

\begin{figure}[hbt!]
\includegraphics[width=0.9\linewidth]{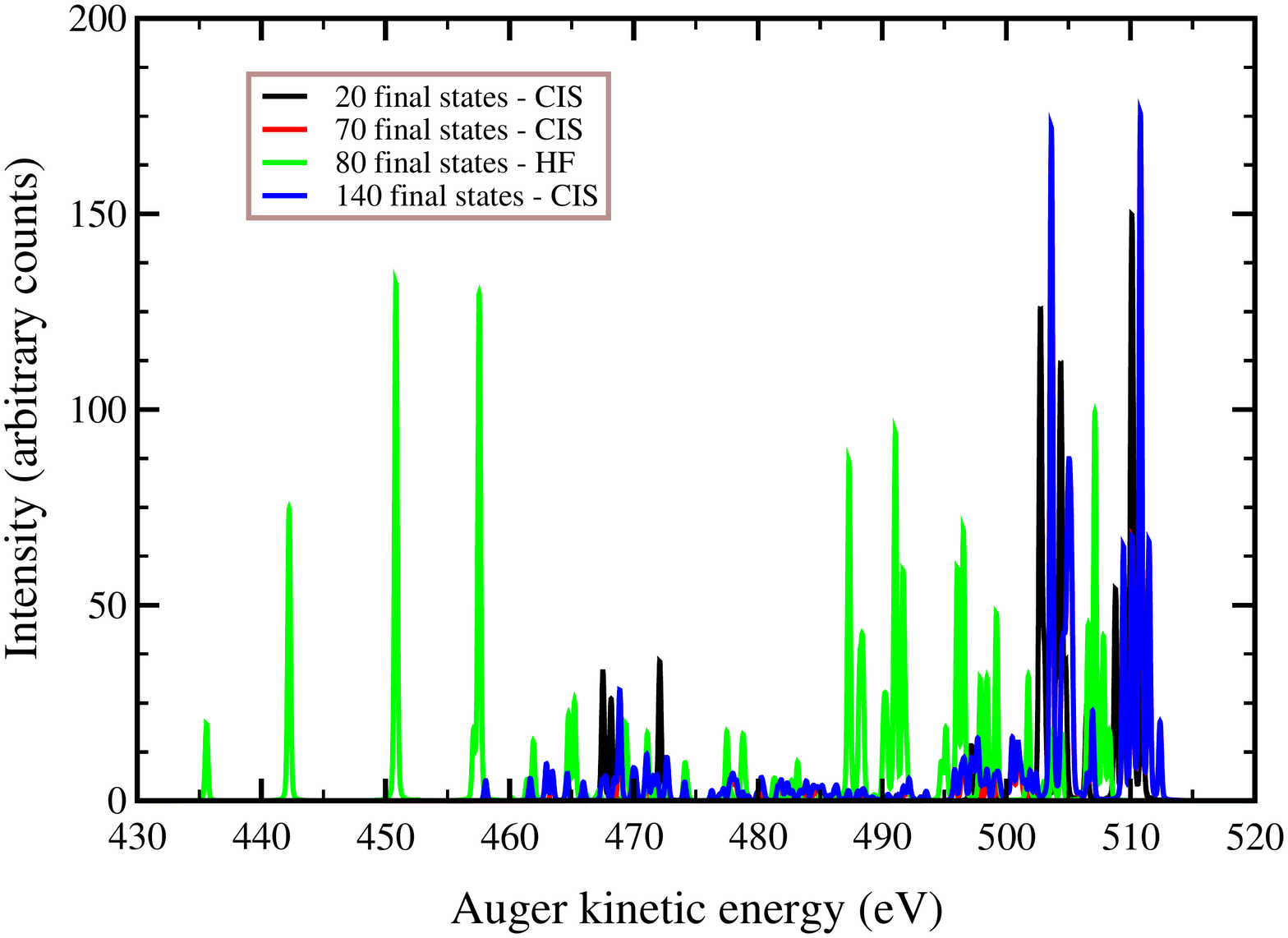}
\caption{$K-LL$ Auger spectrum of ozone, where $K=1b_1$. Green line: HF level of theory with 80 independent final states, obtained by diagonalization of the multichannel Hamiltonian~\eqref{TH11}. Black, red, and blue lines represent the spectral lineshape obtained with an active space of 27 molecular orbitals and a number of 20, 70, 140 independent final channels, respectively. \label{fig3}}
\end{figure}

\begin{figure}[hbt!]
\includegraphics[width=0.9\linewidth]{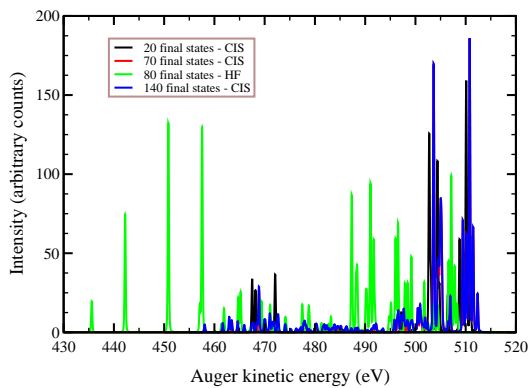}
\caption{$K-LL$ Auger spectrum of ozone, where $K=2a_1$. Green line: HF level of theory with 80 independent final states, obtained by diagonalization of the multichannel Hamiltonian~\eqref{TH11}. Black, red, and blue lines represent the spectral lineshape obtained with an active space of 27 molecular orbitals and a number of 20, 70, 140 independent final channels, respectively. \label{fig4}}
\end{figure}

Finally, in Figure~\eqref{fig2a} we compare our ab-initio calculations with the experimental data of Ref. \cite{doi:10.1063/1.1385156} recorded at a photon energy of 536.7 eV, finding an overall good agreement. We stress that the experimental data refer to a spectator transition $O1s\rightarrow \sigma^*$, which is a process similar (not exactly the same) to that one we considered here. Auger lineshapes were broadened by convolution with a 0.8 eV Gaussian function. These large observed linewidths have been attributed both to the dissociative character of the final states and to nuclear vibrations.

\subsection{The autoionization spectrum of ozone}

In this section we deal with the resonant autoionization spectrum of ozone following the photo-electronic excitation 
$O1s \rightarrow \sigma^*$. These results have been obtained at the ground state geometry, disregarding the nuclear motion. 

The transition energy of this process is 538.39 eV for the two $O_T$ atoms related by $C_{2v}$ symmetry operations and 543.73 eV if the excitation involves the central oxygen atom $O_C$. Indeed, while molecular orbitals were obtained by carrying out CI on HF optimized symmetry orbitals, which are thus delocalized all over ozone, the core-hole almost relocalizes in $O_T$ and $O_C$ upon CI.  

\begin{figure}[hbt!]
\includegraphics[width=0.9\linewidth]{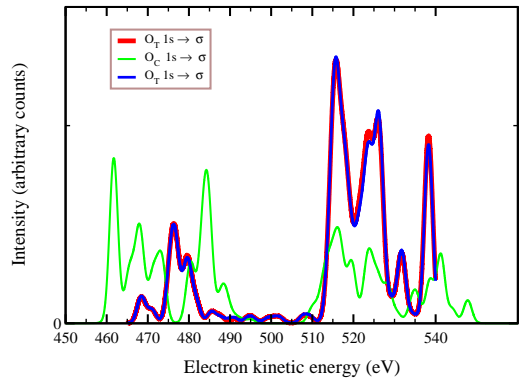}
\caption{$O1s \rightarrow \sigma^*$ autoionization spectrum of ozone. Red, blue lines: CIS lineshapes with 140 final states obtained upon diagonalization of the multichannel Hamiltonian~\eqref{TH11}. Core-hole belongs to molecular symmetry orbitals, which are optimized using HF and extend all over the ozone molecule. However, upon CI the hole relocalizes in one of the two oxygen atoms ($O_T$) related by symmetry operations. Green line: spectral lineshape for a core-hole relocalized after the CIS procedure in the central oxygen atom $O_C$. \label{fig6}}
\end{figure}

In Figure~\eqref{fig6} we plot the behavior of both the resonant and direct contributions to the spectrum as a function of the autoionized electron kinetic energy. Plotted lineshapes have been obtained after convolution with Lorentzian functions, whose width is provided by our ab-initio calculations (see Equation~\eqref{TF10}), and by a further convolution with 1 eV FWHM Gaussian profile to take into account the nuclear broadening and the dissociative nature of the excited states of ozone. In particular, red and blue lines represent CIS calculations of the resonant autoionization spectra of ozone, in which a core-hole is created upon excitation from the molecular symmetry orbital that is almost localized in one of the two oxygen atoms ($O_T$) related by symmetry operations. In green we plot the spectral lineshape for a core-hole created in the symmetry orbital almost localized in the central oxygen atom ($O_C$), whose atomic orbital of $1s$ character (corresponding to the $1a^{-1}$ molecular symmetry orbital) is lower in energy than the other two $1s$ oxygen orbitals ($1b^{-1}, 2a^{-1}$). In these simulations 140 final states have been included. 

\subsection{Core-hole orbital symmetry and localization}

In Auger experiments core-electrons are extracted from a localized inner orbital of one particular oxygen. In our previous calculations core-hole localization was obtained by including single excitations in HF molecular symmetry orbitals. This means that at HF level molecular orbitals were optimized by assuming that the core-hole was ``delocalized'' over the ozone molecule.  
Core-hole localization was achieved basically by adding part of the dynamical correlation neglected by HF via CIS procedure. 

\begin{figure}[hbt!]
\includegraphics[width=0.9\linewidth]{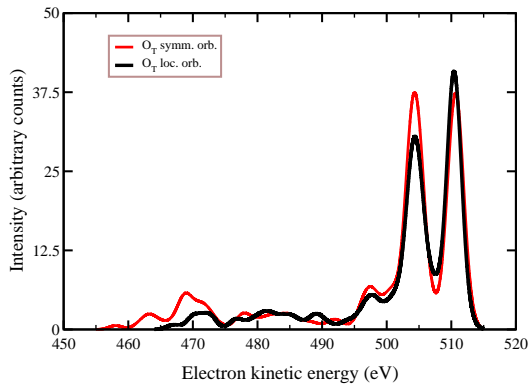}
\caption{$K-LL$ Auger spectrum of ozone obtained from CIS calculations after convolution with Lorentzian functions having the theoretical widths and with 1 eV FWHM Gaussian functions to reproduce the combined effect of the finite resolution of the spectrometer and of the nuclear vibration. Black line: HF orbitals optimized starting from an atomic core-hole localized in on of the two $O_T$ oxygen centers. 
Red line: HF orbitals are optimized with the initial core-hole ``delocalized'' in the molecular symmetry orbitals. Calculations were carried out with an active space of 27 molecular orbitals and 140 independent final channels, including the interchannel coupling~\eqref{TH11}. \label{local}}
\end{figure}

However, it is interesting (and actually more representative of what actually happens in experiments) to show the $K-LL$ Auger spectrum obtained by optimizing the molecular orbitals starting from an atomic core-hole completely localized in one of the two $O_T$s, equivalent by $C_{2v}$ symmetry operation (thus using atomic orbitals in the HF optimization to break the orbital symmetry). 
Indeed the holes, created upon primary
ionization in localized atomic sites, favour intra-atomic more than inter-atomic transitions. This difference in the HF kick-off orbitals reflects into the final spectrum plotted in Figure~\eqref{local} (black line), obtained at the same level of theory as in previous calculations (red line). In particular, the spectral intensity of the highest energy peaks is quite modified by core-hole localization.   

\section{Conclusions}

In this work a first-principles method for the calculation and interpretation of core-electron spectra of ozone has been described.
This method is based on Fano's theory of decay, due to the interaction between discrete and continuum states. We discussed this theoretical framework in the general case of several continua interacting with several quasi-bound
states, including explicitly the boundary conditions appropriate to the
study of Auger emission.

To achieve a relevant reduction in the computational efforts
required by the application of the theory, our approach uses a model Hamiltonian with a projected potential
represented in terms of $L^2$-functions. The spectral properties of this projected Hamiltonian are such that the subspace of the scattering eigenstates significant for the physical problem is finite and isomorphous to the subspace of the functions used for representing the potential
energy operator.  
Furthermore, the many-body problem can be reduced to an effective single-particle problem,
in which the scattering states can be obtained from the resolution of a projected
Lippmann-Schwinger equation with the proper boundary condition.

Wave functions of the bound and continuum orbitals have been expanded onto a $L^2$ Gaussian basis set, which includes decontracted diffuse functions, therefore reducing the Schr{\"o}dinger equation to a matrix equation. 
The calculation of mono and bielectronic integrals (and of the transition matrix elements between discrete and continuum states with non-orthonormal orbitals) is carried out numerically on the basis of analytical expressions.
Furthermore, to assess accurately the partial decay rates into different channels we introduced the effects of interactions among final decay states along with the many-body interaction within the remaining ionic system.

Finally, the analytic expression of the autoionization and Auger cross sections has been applied to predict the ozone radiationless decay lineshape as functions of the kinetic energy of the emitted electron. 
These Auger lineshapes have been broadened with a Lorentzian profile, whose width is a direct result of our ab-initio calculation.
We included also ``a posteriori'' the specific features of the incident
radiation, of the electron spectrometer, and of the effects of nuclear motion in the framework of the Born-Oppenheimer approximation by convolution with a Gaussian profile, finding a good comparison with available experimental data.

The effect of electronic correlation and the influence of core-hole localization on the Auger peak energy positions and intensities have been also analyzed, finding different spectral properties when the HF molecular orbitals are optimized starting from a core-hole localized in a specific oxygen atom that favours intra-atomic transitions. While localization breaks the symmetry of the molecular orbitals, 
which is afterwards recovered by multi-configuration expansion of the wavefunctions, both intensity and energy peak differ than those obtained by delocalizing the inner-hole on the two symmetry-equivalent oxygen centers.  

Our method is applicable to molecular systems owing to the polycentric
nature of the basis functions and to the fact that all the integrals
involved in the calculation of the Auger matrix elements are analytical, cost-effective and easily
programmable. Moreover, it is possible to take into account the effects of
interaction among the final channels without resorting to numerical
integration on the energy as foreseen by Fano's formulae. Finally, the
possibility of obtaining continuum wave functions makes the method
useful also for studying other scattering problems, such as photoionization, internal conversion, electron-atom and electron-molecule scattering and so on.









\newpage
\appendix*
\onecolumngrid
\section{The Auger spectrum of ozone}

\begin{table}[hbt!] 
\caption{Auger spectrum of ozone with a primary hole localized in the $1a_{1}^{-1}$ molecular orbital. First column: double occupancy of the final transition states. Second: Auger transition energy ($E_\alpha$, eV). Third column: partial Auger probabilities ($\Gamma_\alpha$, units: a.u. $\times10^{-3}$) according to Equation~\eqref{TF10}. The theory used is CIS with an active space of 27 orbitals. \label{tab4}}
\begin{tabular}{ccc}
\toprule
\textbf{\footnotesize Trans. state occ.} & \textbf{ \footnotesize Trans. En.}	& \textbf{\footnotesize Trans. Prob.}\\
\hline
 2.00  |   2.00 |  
1.97  |  1.59  |  1.42 |   1.64 |   0.09  |  1.71   &   495.17  & 0.1100 \\ 0.39 |   2.00 |   1.94  |  1.65 |   1.60 |   0.13  |  1.86 & $~$ & $~$  \\
\hline
2.00 |   2.00 |  
1.58 |   1.63 |   1.77 |   1.86 |   0.37 |   1.69   & 467.14  & 0.1818  \\ 0.15 |   2.00  |  1.58 |   1.67 |   1.84  |  0.25 |   1.61 & $~$ & $~$  \\
\hline
2.00 |   2.00 |  
2.00 |   1.95  |  1.63 |   1.53 |   0.19 |   1.38 
 & 502.3 &  0.0879 \\ 0.34  |  2.00 |   1.99 |   1.89 |   1.59 |   0.17  |  1.34 & $~$ & $~$  \\
\hline
2.00 |   2.00 |  
2.00 |   1.96 |   1.70 |   1.62 |   0.14 |   1.39   & 506.25 & 0.0767  \\ 0.54  |  2.00  |  1.99  |  1.90  |  1.68  |  0.05  |  1.01 & $~$ & $~$\\
\hline
2.00 |   2.00 |  
1.99 |   1.94  |  1.80 |   1.53  |  0.34 |   1.21 & 500.02 & 0.0804 \\ 0.21 |   2.00 |   1.99 |   1.89 |   1.46 |   0.24 |   1.40 & $~$ & $~$ \\
\hline
2.00 |   2.00 |  
2.00 |   1.96 |   1.94 |   1.78 |   0.06 |   1.50 & 506.065 & 0.0697  \\ 0.55 |   2.00 |   1.98 |   1.31 |   1.91 |   0.03 |   0.98 & $~$ & $~$ \\
\hline
2.00 |   2.00 |  
1.99 |   1.93 |   1.74  |  1.54 |   0.14  |  1.44  & 506.03 & 0.0761  \\ 0.53 |   2.00 |   1.99 |   1.88 |   1.65 |   0.06 |   1.12  & $~$ & $~$ \\
\hline
2.00 |   2.00 |  
2.00 |   1.88 |   1.67 |   1.49 |   0.21 |   1.90   &  503.195 & 0.0596 \\ 0.18 |   2.00 |   1.98 |   1.33 |   1.70 |   0.12 |   1.55  & $~$ & $~$ \\
\hline
2.00  |  2.00 |  
1.69 |   1.67  |  1.45 |   1.83 |   0.19 |   1.79  & 476.33 & 0.0352 \\ 0.44  |  2.00  |  1.61  |  1.77  |  1.73 |   0.15 |   1.67 & $~$ & $~$\\
\hline
2.00  |  2.00 |  
1.99  |  1.90  |  1.47  |  1.35 |   0.17  |  1.81   & 502.25 & 0.0693 \\ 0.26 |   2.00 |   1.97 |   1.73 |   1.50 |   0.27  |  1.58 & $~$ & $~$\\
\hline
2.00 |   2.00 |  
2.00 |   1.98 |   1.36 |   1.22 |   0.03 |   1.95  & 508.36 & 0.0705   \\ 0.30 |   2.00 |   1.99 |   1.82 |   1.59 |   0.03 |   1.74 & $~$ & $~$\\
\hline
2.00 |   2.00 |  
2.00 |   1.96 |   1.46 |   1.52 |   0.10 |   1.97  &   504.16  & 0.0518\\ 0.54 |   2.00  |  1.99  |  1.67 |   1.21 |   0.11 |   1.47  & $~$ & $~$\\
\hline
2.00 |   2.00 |  
2.00 |   1.98 |   1.94 |   1.13 |   0.02 |   1.54   &  512.03 & 0.0553\\ 0.59 |   2.00 |   2.00 |   1.98 |   1.93 |   0.01 |   0.89 & $~$ & $~$\\
\hline
 2.00 |   2.00 |  
1.34  |  1.59 |   1.78 |   1.86 |   0.30 |   1.87   & 453.39 & 0.1166  \\ 0.29 |   2.00 |   1.41 |   1.76 |   1.72 |   0.27 |   1.81 & $~$ & $~$ \\
\hline
 2.00 |   2.00 |  
1.99  |  1.73  |  1.67  |  1.59 |   0.09 |   1.77  & 503.62 & 0.0466   \\0.44 |   2.00 |   1.97  |  1.68  |  1.25 |   0.05 |   1.77  & $~$ & $~$ \\
\hline
 2.00 |   2.00 |  
1.99  |  1.87  |  1.80  |  1.67  |  0.28  |  1.42  & 496.73 & 0.0512  \\ 0.28 |   2.00  |  1.99  |  1.63 |   1.50  |  0.26 |   1.31 & $~$ & $~$ \\
\hline
2.00 |   2.00 |  
2.00  |  1.95  |  1.95 |   1.49 |   0.49 |   1.83  &  502.45 & 0.0416 \\ 0.16 |   2.00  |  1.99  |  1.35 |   1.84 |   0.06 |   0.90 & $~$ & $~$ \\
\hline
 2.00  |  2.00 |  
1.99  |  1.82 |   1.81 |   1.67  |  0.21 |   1.61  & 500.80 & 0.0432 \\ 0.23 |   2.00 |   1.99 |   1.60  |  1.59 |   0.27 |   1.21 & $~$ & $~$ \\
\hline
\textbf{Total Transition Probability ($\Gamma$)} & $1.4321 \times 10^{-3}$ A.U. & $~$  \\ 
\end{tabular}
\end{table}

\begin{table}[hbt!] 
\caption{Auger spectrum of ozone with a primary hole localized in the $1b_{1}^{-1}$ molecular orbital. First column: double occupancy of the final transition states. Second: Auger transition energy ($E_\alpha$, eV). Third column: partial Auger probabilities ($\Gamma_\alpha$, units: a.u. $\times10^{-3}$) according to Equation~\eqref{TF10}. The theory used is CIS with an active space of 27 orbitals. \label{tab5}}
\begin{tabular}{ccc}
\toprule
\textbf{\footnotesize Trans. state occ.} & \textbf{ \footnotesize Trans. En.}	& \textbf{\footnotesize Trans. Prob.}\\
\hline
2.00  |  2.00 |  
2.00  |  1.99 |   1.98  |  1.92  |  0.04  |  1.93  & 510.751 & 0.2255  \\ 0.03 |   2.00  |  1.99 |   1.98 |   1.87  |  0.03  |  0.24  & $~$ & $~$  \\
\hline
2.00 |   2.00 |  
2.00  |  2.00  |  1.97  |  1.90  |  0.03 |   1.86 
  & 510.17 & 0.1713 \\ 0.08  |  2.00 |   1.99  |  1.98  |  1.10  |  0.02  |  1.07 & $~$ & $~$  \\
\hline
2.00 |   2.00 |  
2.00  |  1.98  |  1.46  |  1.58  |  0.05  |  1.89  & 502.71 & 0.1755\\ 0.38  |  2.00  |  1.98  |  1.80 |   1.19 |   0.03  |  1.67  & $~$ & $~$  \\
\hline
2.00 |   2.00 |  
2.00  |  1.98 |   1.36  |  1.22  |  0.03  |  1.95  &  502.79 & 0.1453 \\ 0.30 |   2.00 |   1.99 |   1.82  |  1.59  |  0.03  |  1.74 & $~$ & $~$\\
\hline
2.00 |   2.00 |  
2.00 |   1.97 |   1.49 |   1.94 |   0.03 |   1.78 
  &   504.44 & 0.1286 \\ 0.33  |  2.00 |   1.99 |   1.98 |   1.56  |  0.02 |   0.90  & $~$ & $~$ \\
\hline
2.00 |   2.00 |  
2.00  |  1.99 |   1.97  |  1.09  |  0.02  |  1.90  & 510.93 & 0.1439  \\ 0.07  |  2.00  |  2.00 |   1.98  |  1.94  |  0.02 |   1.03 & $~$ & $~$ \\
\hline
2.00  |  2.00 |  
2.00  |  1.97  |  1.84  |  1.29  |  0.02 |   1.95   & 508.79 & 0.1289  \\  0.18 |   2.00  |  1.99  |  1.96 |   1.09 |   0.02  |  1.69 & $~$ & $~$ \\
\hline
2.00  |  2.00 |  
2.00  |  1.97  |  1.84  |  1.03  |  0.02  |  1.98  & 509.47 & 0.1359 \\ 0.19 |   2.00  |  2.00 |   1.97 |   1.16 |   0.01 |   1.82  & $~$ & $~$ \\
\hline
2.00  |  2.00 |  
2.00  |  1.99  |  1.98 |   1.81 |   0.04  |  1.10  & 504.35 &  0.1162  \\ 0.17 |   2.00 |   1.99  |  1.88 |   1.93 |   0.04 |   1.09  & $~$ & $~$ \\
\hline
2.00 |   2.00 |  
2.00  |  1.98 |  1.91  |  1.67  |  0.07 |   1.62   & 503.07 & 0.0798  \\ 0.33  |  2.00 |   1.99 |   1.52 |   1.82  |  0.03  |  1.07  & $~$ & $~$ \\
\hline
2.00 |   2.00 |  
2.00 |   1.92  |  1.87 |   1.69 |   0.02 |   1.92  &  503.99 & 0.0706   \\ 0.25 |   2.00 |   1.98  |  1.32 |   1.20  |  0.01  |  1.83  & $~$ & $~$\\
\hline
2.00 |   2.00 |  
2.00 |   1.98 |   1.94 |   1.67 |   0.02  |  1.68   & 504.29 & 0.0608 \\ 0.29  |  2.00 |   1.98 |   1.52  |  1.86 |   0.01  |  1.06 & $~$ & $~$\\
\hline
2.00 |   2.00 |  
2.00  |  1.98  |  1.94  |  1.13 |   0.02 |   1.54 & 506.46  & 0.0554  \\ 0.59 |   2.00 |   2.00 |   1.98 |   1.93 |   0.01 |   0.89   & $~$ & $~$\\
\hline
2.00  |  2.00 |  
2.00  |  1.94  |  1.79 |   1.10  |  0.01  |  1.81  & 504.77 & 0.0777  \\ 0.47  |  2.00 |   1.99  |  1.59  |  1.65  |  0.01 |   1.64  & $~$ & $~$\\
\hline
2.00  |  2.00 |  
1.91  |  1.62 |   1.55  |  1.65 |   0.29 |   1.83   &  472.09 & 0.0856 \\ 0.32 |   2.00 |   1.65  |  1.50  |  1.65 |   0.22 |   1.80 & $~$ & $~$ \\
\hline
 2.00 |   2.00 |  
1.99 |   1.84 |   1.77  |  1.16  |  0.42 |   1.95  & 497.19 & 0.0339 \\ 0.28 |   2.00 |   1.96 |   1.59 |   1.23 |   0.04 |   1.76 & $~$ & $~$ \\
\hline
2.00  |  2.00 |  
1.73  |  1.70 |   1.70 |   1.57 |   0.34 |   1.91  & 467.51 & 0.0790\\ 0.15  |  2.00  |  1.62  |  1.44  |  1.62 |   0.30  |  1.90 & $~$ & $~$ \\
\hline
2.00 |   2.00 |   1.99  |  1.60
1.88  |  1.54  |  0.25  |  1.67  & 499.45 & 0.0032 \\ 0.18  |  2.00  |  1.99  |  1.83 |   1.87 |   0.05 |   1.15 & $~$ & $~$ \\
\hline
2.00 |   2.00 |  
1.68 |   1.70 |   1.68  |  1.68 |   0.19  |  1.68  & 468.17 & 0.0622  \\ 0.51 |   2.00 |   1.70 |   1.70  |  1.53  |  0.18 |   1.78  & $~$ & $~$ \\
\hline
 2.00 |   2.00 |   1.99  |  1.88 |  
1.86  |  1.81 |   0.22  |  1.54  & 491.23 & 0.0032 \\ 0.37  |  2.00 |   1.73 |   1.62 |   1.74 |   0.14  |  1.09  & $~$ & $~$\\
\hline
\textbf{Total Transition Probability ($\Gamma$)} & $1.9824 \times 10^{-3}$ A.U. & $~$  \\ 
\end{tabular}
\end{table}

\begin{table}[hbt!] 
\caption{Auger spectrum of ozone with a primary hole localized in the $2a_{1}^{-1}$ molecular orbital. First column: double occupancy of the final transition states. Second: Auger transition energy ($E_\alpha$, eV). Third column: partial Auger probabilities ($\Gamma_\alpha$, units: a.u. $\times10^{-3}$) according to Equation~\eqref{TF10}. Last row: total decay rate. The theory used is CIS with an active space of 27 orbitals. \label{tab6}}
\begin{tabular}{ccc}
\toprule
\textbf{\footnotesize Trans. state occ.} & \textbf{ \footnotesize Trans. En.}	& \textbf{\footnotesize Trans. Prob.}\\
\hline
 2.00 | 2.00 |
2.00 | 1.99 | 1.98 | 1.92 | 0.04 | 1.93  &  510.07  & 0.2515 \\ 0.03 | 2.00 | 1.99 | 1.98 | 1.87 | 0.03 | 0.24 & $~$ & $~$  \\
\hline
2.00 | 2.00 |
2.00 | 2.00 | 1.97 | 1.90 | 0.03 | 1.86  & 510.17 & 0.1693 \\ | 0.08 | 2.00 | 1.99 | 1.98 | 1.10 | 0.02 | 1.07 & $~$ & $~$  \\
\hline
2.00 | 2.00
2.00 | 1.98 | 1.46 | 1.58 | 0.05 | 1.89  & 502.71 & 0.1708 \\ 0.38 | 2.00 | 1.98 | 1.80 | 1.19 | 0.03 | 1.67 & $~$ & $~$  \\
\hline
2.00 | 2.00 |
2.00 | 1.98 | 1.36 | 1.22 | 0.03 | 1.95  &  502.79 & 0.1478 \\ 0.30 | 2.00 | 1.99 | 1.82 | 1.59 | 0.03 | 1.74 & $~$ & $~$\\
\hline
2.00 | 2.00 |
2.00 | 1.97 | 1.49 | 1.94 | 0.03 | 1.78  &  504.44  & 0.1290 \\ 0.33 | 2.00 | 1.99 | 1.98 | 1.56 | 0.02 | 0.90  & $~$ & $~$ \\
\hline
2.00 | 2.00 |
2.00 | 1.97 | 1.84 | 1.29 | 0.02 | 1.95  & 508.79 & 0.1401 \\ 0.18 | 2.00 | 1.99 | 1.96 | 1.09 | 0.02 |  1.69 & $~$ & $~$ \\
\hline
2.00 | 2.00
2.00 | 1.99 | 1.97 | 1.09 | 0.02 | 1.90  & 510.92 & 0.1443 \\  0.07 | 2.00 | 2.00 | 1.98 | 1.94 | 0.02 | 1.03  & $~$ & $~$ \\
\hline
2.00 |  2.00
2.00  | 1.97 |  1.84  | 1.03  | 0.02 |  1.98 & 509.46 & 0.1225 \\ 0.19 |  2.00  | 2.00 |  1.97 |  1.16 |  0.01  | 1.82  & $~$ & $~$ \\
\hline
2.00 |  2.00 | 
2.00  | 1.99  | 1.98  | 1.81 |  0.04 |  1.10 &  504.35 & 0.1091 \\ 0.17 |  2.00 |  1.99 |  1.88 |  1.93 |  0.04  | 1.09  & $~$ & $~$ \\
\hline
2.00 |  2.00 | 
2.00 |  1.98 |  1.91 |  1.67 |  0.07 |  1.62  & 503.06 & 0.0797 \\ 0.33 |  2.00 | 1.99 |  1.52  | 1.82 |  0.03 |  1.07  & $~$ & $~$ \\
\hline
 2.00 |  2.00 | 
2.00  | 1.92 |  1.87 |  1.69  | 0.02 |  1.92 & 503.99 & 0.0808 \\ 0.25 2.00  | 1.98  | 1.32 |  1.20 |  0.01 |  1.83  & $~$ & $~$\\
\hline
2.00  | 2.00 | 
2.00  | 1.98 |  1.94  | 1.13 |  0.02  | 1.54  & 506.45 & 0.0552 \\ 0.59 |  2.00 |  2.00 |  1.98 |  1.93 |  0.01 |  0.89& $~$ & $~$\\
\hline
2.00 |  2.00 | 
2.00 |  1.98 |  1.94 |  1.67 |  0.02 |  1.68 & 504.28 & 0.0610 \\ 0.29 |  2.00 |  1.98 |  1.52 |  1.86 |  0.01 |  1.06  & $~$ & $~$\\
\hline
2.00  |  2.00 |  
2.00  |  1.94 |   1.79 |   1.10  |  0.01  |  1.81 & 504.77 & 0.0663 \\ 0.47  |  2.00 |   1.99  |  1.59  |  1.65  |  0.01 |   1.64  & $~$ & $~$\\
\hline
 2.00 |  2.00 | 
1.91 |  1.62 |  1.55  | 1.65 |  0.29 |  1.83 & 472.09 &  0.0878 \\ 0.32  | 2.00 |  1.65 |  1.50 |  1.65 |  0.22 |  1.80  & $~$ & $~$\\
\hline
2.00  | 2.00 | 
1.99 |  1.84 |  1.77 |  1.16 |  0.42  | 1.95   & 497.19 & 0.030\\ 0.28 |  2.00 | 1.96  | 1.59  | 1.23  | 0.04 |  1.76 & $~$ & $~$ \\
\hline
 2.00 |  2.00  | 1.99 |  1.60 | 
1.88  | 1.54 |  0.25 |  1.67 & 499.45 & 0.0033 \\ 0.18 |  2.00 |  1.99  | 1.83  | 1.87 |  0.05 |  1.15  & $~$ & $~$ \\
\hline
2.00 |  2.00 | 
1.73  | 1.70 |  1.70 |  1.57 |  0.34 |  1.91 & 467.51 & 0.0799  \\ 0.15  | 2.00  | 1.62 |  1.44  | 1.62  | 0.30 |  1.90 & $~$ & $~$ \\
\hline
2.00 |  2.00 | 
1.68 |  1.70  | 1.68 |  1.68  | 0.19 |  1.68 & 468.17 & 0.0637 \\ 0.51 |  2.00  | 1.70 |  1.70  | 1.53 |  0.18 |  1.78 & $~$ & $~$ \\
\hline
2.00 |  2.00 |  1.99  | 1.88 | 
1.86  | 1.81  | 0.22  | 1.54 & 491.23 & 0.0032 \\ 0.37 |  2.00 |  1.73 |  1.62 |  1.74  | 0.14 |  1.09  & $~$ & $~$ \\
\hline
\textbf{Total Transition Probability ($\Gamma$)} & $1.9957 \times 10^{-3}$ A.U. & $~$  \\ 
\end{tabular}
\end{table}

\end{document}